\documentclass[journal]{IEEEtran}
\usepackage{setspace}
\usepackage{fancyhdr}
\usepackage{graphicx}
\usepackage{color}
\usepackage{placeins}
\usepackage{float}
\usepackage{tabularx,colortbl}
\usepackage{amsmath}
\usepackage{amsfonts}
\usepackage{amssymb}
\usepackage{amsthm}
\usepackage[numbers,sort&compress]{natbib}
\usepackage[linesnumbered,boxed,ruled,commentsnumbered]{algorithm2e}
\usepackage{algorithmic}
\usepackage{bm}
\usepackage{epstopdf}
\usepackage{mathrsfs}
\usepackage{multirow}
\usepackage{xcolor}

\begin{document}
%
\title{Learning Oriented Cross-Entropy Approach to User Association in Load-Balanced HetNet}
\author{
\IEEEauthorblockN{Xietian~Huang, Wei~Xu, Guo~Xie, Shi~Jin, and Xiaohu~You}
\vspace{-0.8cm}
\thanks{
Manuscript received April 06, 2018; accepted June 05, 2018. This work of was supported in part by NSFC (grant nos. 61471114, 61625106, 61571118, U1534208) and  the Six Talent Peaks project in Jiangsu Province under GDZB-005. The editor coordinating the review of this paper and approving it for publication was M. Kountouris. \textit{(Corresponding author: Wei Xu.)}

X. Huang, W. Xu, S. Jin, and X. You are with the National Mobile Communications Research Laboratory (NCRL), Southeast University, Nanjing, China (\{xthuang, wxu, jinshi, xhyu\}@seu.edu.cn).

G. Xie is with the Shaanxi Key Laboratory of Complex System Control and Intelligent Information Processing, Xi'an University of Technology, Xi'an, China (guoxie@xaut.edu.cn).
}

}



%


\maketitle

\newtheorem{mylemma}{Lemma}
\newtheorem{mytheorem}{Theorem}
\newtheorem{mypro}{Proposition}
\begin{abstract}
This letter considers optimizing user association in a heterogeneous network via utility maximization, which is a combinatorial optimization problem due to integer constraints. Different from existing solutions based on convex optimization, we alternatively propose a cross-entropy (CE)-based algorithm inspired by a sampling approach developed in machine learning. Adopting a probabilistic model, we first reformulate the original problem as a CE minimization problem which aims to learn the probability distribution of variables in the optimal association. An efficient solution by stochastic sampling is introduced to solve the learning problem. The integer constraint is directly handled by the proposed algorithm, which is robust to network deployment and algorithm parameter choices. Simulations verify that the proposed CE approach achieves near-optimal performance quite efficiently.
\end{abstract}

\vspace{-0.4cm}

%
\IEEEpeerreviewmaketitle
\section{Introduction}
\vspace{-0.1cm}
Recently, heterogeneous network (HetNet) has become both an academic and an industrial focus due to its advantage in enhancing spectral efficiency. HetNet layout equips many low-power small-cell base stations (SBSs) overlaid with macro-cell base station (MBS), which helps to promote the network performance while also coming with multiple challenges.

User association is one of the issues that need to be reconsidered in HetNet \cite{Yang2017User}-\cite{Shen2014Distributed}. Because the transmit power of MBS is much higher than SBS, most users may stick to their MBS association based on received signal strength, which leads to unbalanced load. To make the best use of the heterogeneous infrastructure, users should be transferred to lightly loaded SBSs. In this way, users are better served with more available resource and the entire network benefits. Consequently, balanced user associations are essential in reaping the benefits of HetNet. Most existing literature, e.g., \cite{Yang2017User}-\cite{Huang2018Utility}, solved the association problems resorting to convex optimization including the commonly adopted Lagrangian dual decomposition and subgradient methods. These methods, however, are sensitive to algorithm parameters \cite{Shen2014Distributed}, which hinders wide applications in practice.

In this letter, we investigate the load-balancing association problem from the machine learning perspective in that the optimal association is regarded as a random variable whose probability distribution is designed to be dynamically learned via efficient stochastic sampling. The CE approach was firstly introduced in 1997 \cite{Rubinstein2004The} and developed in machine learning. The advantage of the CE approach lies in its adaptive update procedure \cite{Kovaleva2017Cross}, which makes it be inherently capable of solving combinatorial optimization problems in a much simpler way than typical relaxation techniques. To the best of our knowledge, this is the first time that CE method is used to solve the constrained user association problem. Specifically, the proposed approach first randomly generates candidate association matrices and chooses some elites by evaluating the objective values. By refining the probability distributions iteratively via CE minimization, this approach yields a near-optimal association with sufficiently high probability. Compared to existing methods, our proposed algorithm is more robust, i.e., nonsensitive, to network deployment and algorithm parameter choices. Besides, simulation results verify that the proposed algorithm achieves near-optimal performance in terms of both utility rate and load balancing.

\section{System Model}\label{sec:system_model}
\vspace{-0.1cm}
We consider a typical downlink HetNet consisting of $I$ users and $J$ BSs including MBSs and SBSs. Let $\mathcal{I}\!=\!\{1,2,...,I\}$ and $\mathcal{J}\!=\!\{1,2,...,J\}$ denote the sets of users and BSs, respectively. The received signal-to-interference-plus-noise ratio (SINR) is
\begin{equation}\label{eq:SINR}
 \text{SINR}_{ij}=\frac{h_{ij}P_{j}}{\sum_{q\neq j}h_{iq}P_{q}+\sigma^{2} },\; \forall i\in \mathcal{I}, j\in \mathcal{J}
\end{equation}
where $h_{ij}$ denotes the channel gain between user $i$ and BS $j$, $P_{j}$ is the transmit power of BS $j$, and $\sigma^{2}$ is the noise power.

Denote binary variables $\{x_{ij}\}$ as the indicator of the association between user $i$ and BS $j$. If user $i$ is associated with BS $j$, then $x_{ij}\!=\!1$, otherwise $x_{ij}\!=\!0$.
Let $W$ denote the system bandwidth, which is reused by all BSs. Users associated with the same BS share the frequency resource. Assuming a uniform resource allocation among users, the achievable rate $R_{ij}$ can thus be evaluated as
\begin{equation}\label{eq:rate}
 R_{ij}=\frac{W}{\sum_{i\in \mathcal{I}} x_{ij}}\log(1+\text{SINR}_{ij}),\; \forall i\in \mathcal{I}, j\in \mathcal{J}.
\end{equation}
And we obtain the overall rate of user $i$ as $R_i=\sum_{j\in \mathcal{J}}x_{ij}R_{ij}$.

Considering that network utility is the ultimate goal in providing wireless services, we investigate a utility maximization problem in the network via association optimization. Let $U_{i}(R_{i})$ be the utility function of user $i$. Multiple typical utility functions apply depending on the adopted metrics. Specially, adopting an identity function returns to the common rate maximization problem. To achieve load balancing and maximize total user QoS satisfactory, we choose the typical logarithmic utility function, by which QoS requirements of users are guaranteed in the sense of fairness. However, our proposed algorithm is a general approach without any preference on the utility function.

Now, we can formulate the utility optimization problem as:
\begin{subequations}\label{eq:problem}
\begin{align}
\mathop{\max}_{\bm{x}}\quad\!\!
&\sum\limits_{i\in \mathcal{I}}U_{i}\left(\sum\limits_{j\in \mathcal{J}} x_{ij}R_{ij}\right)\\
\textrm{s.t.}\quad\!\!
&\sum\limits_{j\in \mathcal{J}} x_{ij} = 1, \; \forall i\in \mathcal{I}\\
& x_{ij}\in \{0,1\}, \; \forall i\in \mathcal{I}, j\in \mathcal{J}\\
&\sum\limits_{i\in \mathcal{I}} x_{ij} \leq L_j, \; \forall j\in \mathcal{J}
\end{align}
\end{subequations}
where $\bm{x}\!=\!(x_{11},...,x_{1J},...,x_{IJ})^\text{T}$ is the user association vector and $L_j$ is an upper bound which is irrelevant to variable $x_{ij}$. Constraint (3b) denotes that each user is associated with a single BS at a time. Note that constraint (3d) intuitively represents the bound of BS load. By assigning different values to $L_j$, it can also be regard as several practical constraints, e.g., energy constraint in \cite{Han2016A}\cite{Han2017Backhaul}.

\section{Cross-entropy Approach to Association}
\vspace{-0.1cm}
It is worth pointing out that due to the binary constraints of $x_{ij}$, problem (\ref{eq:problem}) is a combinatorial optimization problem, which is in general NP-hard \cite{Liu2016User}. A popular method of circumventing this difficulty is to make the problem convex by relaxing $\{x_{ij}\}$ from $\{0,1\}$ to continuous in $[0,1]$, and then solve the relaxed problem using convex optimization tools. Its optimality, however, may not be preserved for the original problem in theory. Besides, most existing solutions based on convex optimization, e.g., Lagrangian dual decomposition and subgradient methods, are sensitive to algorithm parameters, which implies that they can be less efficient in practice.

Obviously, the global optimum of problem (\ref{eq:problem}) can be obtained by a direct exhaustive search, which has unbearably computation complexity even in a modest-sized HetNet. To solve the nonconvex problem in (\ref{eq:problem}), we reformulate the problem as a probability learning problem and propose a CE-based solution \cite{Rubinstein2013The} with the assist of stochastic sampling. According to our analysis and simulations, the proposed algorithm requires significantly low complexity compared to the exhaustive search while achieves near-optimal performance with comparable complexity of existing convex optimization methods. The approach applies to all utility functions and behaves nonsensitively to parameter choices. It can thus be applied as a competitive alternative to many, not necessarily convex, user association problems.

\subsection{Problem Formulation of Association Learning}
In machine learning field, problems are usually modelled as probability distribution function (PDF) learning procedure to find the best distribution that matches the input-output relationship in training data set. CE approach is a probabilistic model-based method to solve the learning problem in an iterative mechanism. In the user association problem in (\ref{eq:problem}), the aim is to find the optimal $\bm{x}$ maximizing the network utility. Alternatively, we can model the association vector as a random variable, $\bm{x}$, and then the original problem can be regarded as learning the optimal distribution of variable $\bm{x}$. In statistics, the probability of each possible value that a discrete random variable, $\bm{x}$, can take is described by the PDF, $p(\bm{x})$.

In order to obtain the distribution of the near-optimal association, a straightforward way is to use crude Monte-Carlo simulation. First generate random samples and select some
samples which perform well. Assume the PDF of those observed samples as $q(\bm{x})$. Technically, $q(\bm{x})$ can be regarded as an observation from the true distribution. Then solving the problem is equivalent to learning a PDF $p(\bm{x})$ such that the mismatch between the two distributions $q(\bm{x})$ and $p(\bm{x})$ is minimized, i.e., $p(\bm{x})$ can best describe the PDF of those observed well-performing samples. The association vector $\bm{x}$ is thus obtained from the learned $p(\bm{x})$, which can yield maximal network utility with high probability.

Fundamentally, CE is used as an effective measure quantitatively characterizing the difference between two distributions. For discrete random variable $\bm{x}$, it is defined as following:
\vspace{-0.1cm}
\begin{equation}\label{CE}
\mathcal{D}(q,p)\!=\!\mathbb{E}_q \left[\ln\frac{q(\bm{x})}{p(\bm{x})}\right]\!=\!\sum q(\bm{x})\ln q(\bm{x})\!-\!\sum q(\bm{x})\ln p(\bm{x}).
\vspace{-0.1cm}
\end{equation}
Thus, we can model the PDF learning reformulation of (\ref{eq:problem}) as a CE minimization problem which yields
\vspace{-0.1cm}
\begin{equation}\label{eq:minimize_CE1}
\min\limits_{p} \sum q(\bm{x})\ln q(\bm{x})-\sum q(\bm{x})\ln p(\bm{x})
\vspace{-0.1cm}
\end{equation}
under the same constraints of $\bm{x}$ in problem (3). Since the first term $\sum q(\bm{x})\ln q(\bm{x})$ is constant with respect to the desired $p(\bm{x})$, we have the equivalent maximization problem as
\vspace{-0.1cm}
\begin{equation}\label{eq:minimize_CE2}
\max\limits_{p} \sum q(\bm{x})\ln p(\bm{x}), \;\textrm{s.t.}\;(3\text{b})-(3\text{d}).
\vspace{-0.1cm}
\end{equation}

The CE approach is defined to search over the space of all valid distribution functions to find the optimal distribution, which is infeasible in practice. In machine learning, however, a typical way is to restrict $p(\bm{x})$ in one of the popularly used PDF families, which reduces the search procedure from the entire function space to a finite-dimensional variable space. The choice of a probability distribution family in the CE approach strongly depends on the nature of the design variables. For optimization problems with discrete variables, such discrete probability distribution families as Poisson, Bernoulli and discrete uniform can be applied \cite{MacKay2003Information}. For binary association variable $\bm{x}$ in our work, we adopt the Bernoulli distribution which has the PDF given by $p(\bm{x}; \bm{u})$, where $\bm{u}$ is the parameter vector denoting the success probability. Hence, optimization problem (\ref{eq:minimize_CE2}) can be reformulated as:
\begin{equation}\label{eq:minimize_CE3}
\max\limits_{\bm{u}} \sum q(\bm{x})\ln {p(\bm{x};\,\bm{u})}, \;\textrm{s.t.}\;(3\text{b})-(3\text{d}).
\end{equation}

To solve (\ref{eq:minimize_CE3}), we introduce an efficient stochastic sampling method. Specifically, the algorithm first generates $S$ random samples, e.g., feasible association vectors $\bm{x}$ in our problem, according to assumed probability distribution. For each sample of $\{\bm{x}^s\}_{s=1}^{S}$, it simply appears with probability $1/S$, i.e., $q(\bm{x}^s)\!=\!1/S$. Then we compute the objective value, e.g., the sum utility rate in our problem, of each sample and select $S_\text{elite}$ best samples as ``elites''. Therefore, the probability distribution parameter is obtained based on the selected elites by minimizing the CE, which is evaluated as:
\vspace{-0.1cm}
\begin{equation}\label{eq:minimize_CE4}
\bm{u}^{*}=\arg\;\max\limits_{\bm{u}} \frac{1}{S} \sum_{s=1}^{S_\text{elite}} \ln p(\bm{x}^{[s]};\bm{u})
\vspace{-0.1cm}
\end{equation}
where $\bm{x}^{[s]}$ is the association corresponding to the $s{\text{th}}$ item in the resorted sequence obtained from step 5 in \textbf{Algorithm 1}. By following the procedure in each iteration, as inspired in \cite{Kovaleva2017Cross}, the CE approach can produce a sequence of sampling distributions that are increasingly concentrated around the optimal design.

\subsection{The Proposed CE-based Association Algorithm}
\vspace{-0.1cm}
Applying the reformulation in (\ref{eq:minimize_CE3}) and the sampling algorithm in (\ref{eq:minimize_CE4}), we propose the CE-based ASsociation (CEAS) algorithm, which is summarized in \textbf{Algorithm 1}. Vectorize the association variable as $\bm{x}\!=\!(x_1,...,x_N)^\text{T}$, where $N\!=\!IJ$. We set the probability parameter vector as $\bm{u}\!=\!(u_1,...,u_N)^\text{T}$, where $u_n$ ($0\leq u_n\leq1$) denotes the probability of $x_n\!=\!1$. Initially, we assume that all the elements of $\bm{x}$ belong to $\{0, 1\}$ with equal probability since no prior distribution information is available. That is we initialize the probability parameter as $\bm{u}^{(0)}\!=\!\frac{1}{2}\times{\textbf{1}_{N\times 1}}$, where $\textbf{1}$ is the all-one vector.

During the $t$th iteration, we first generate $S$ candidate association vectors $\{\bm{x}^s\}_{s=1}^{S}$ by the stochastic sampling according to the probability $p(\bm{x}^s;\bm{u}^{(t)})$. Note that we simply discard the generated vector that not satisfying constraints (3b) and (3d). Since each sample is a Bernoulli random variable following $\bm{x}^s\sim \text{Ber}(\bm{u}^{(t)})$, the probability of $\bm{x}^s$ is calculated as:
\vspace{-0.1cm}
\begin{equation}\label{eq:pdf}
p(\bm{x}^s;\bm{u}^{(t)})=\prod_{n=1}^{N} (u_n^{(t)})^{x_n^{s}}(1-u_n^{(t)})^{(1-x_n^{s})}
\vspace{-0.1cm}
\end{equation}
which is used in step 3 of \textbf{Algorithm 1}. Then, in step 4, we calculate the objective function $\{F(\bm{x}^s)\}_{s=1}^{S}$, where $F(\bm{x}^s)=\sum\limits_{i\in \mathcal{I}}U_i(\sum\limits_{j\in \mathcal{J}} x_{ij}^{s}R_{ij})$. Sort $\{F(\bm{x}^s)\}_{s=1}^{S}$ in descending order and the elites are obtained in step 6.

The next step is using elites to update $\bm{u}^{(t+1)}$ by minimizing CE, i.e., solving problem (\ref{eq:minimize_CE4}). By substituting (\ref{eq:pdf}) in (\ref{eq:minimize_CE4}), we derive the first-order derivative of the object function in (\ref{eq:minimize_CE4}) with respect to $u_n$ as
\vspace{-0.1cm}
\begin{equation}\label{eq:derivative}
\frac{1}{S} \sum_{s=1}^{S_\text{elite}}\left(\frac{x_n^{[s]}}{u_n}-\frac{1-x_n^{[s]}}{1-u_n}\right).
\vspace{-0.1cm}
\end{equation}
Forcing (\ref{eq:derivative}) to zero, the optimal $u_n$ is obtained as
\vspace{-0.1cm}
\begin{equation}\label{eq:origin_update}
u_n^{*}=\frac{1}{S_\text{elite}}\sum_{s=1}^{S_\text{elite}} x_n^{[s]}.
\vspace{-0.1cm}
\end{equation}

When updating the parameter vector, we use a smoothed updating procedure, which is especially relevant for CE-based approaches involving discrete random variables \cite{Boer2005A}. The probability parameter is in practice updated by
\vspace{-0.1cm}
\begin{equation}\label{eq:smooth_update}
 \bm{u}^{(t+1)}=\alpha \bm{v}^{(t)}+ (1-\alpha)\bm{u}^{(t)}
\vspace{-0.1cm}
\end{equation}
where $\bm{v}^{(t)}$ is the vector obtained via (\ref{eq:origin_update}) and $\alpha$ is a factor satisfying $0\leq \alpha \leq 1$.

\IncMargin{1em}
\begin{algorithm}

    \textbf{Initialize:} $t=0$; $\bm{u}^{(0)}=\frac{1}{2}\times{\textbf{1}_{N\times 1}}$.\\
    \textbf{for} $t=0:T$ \\
    Generate feasible $\{\bm{x}^s\}_{s=1}^{S}$ based on $p(\bm{x}^s;\bm{u}^{(t)})$\;

    Calculate the objective function $\{F(\bm{x}^s)\}_{s=1}^{S}$\;

    Sort $\{F(\bm{x}^s)\}_{s=1}^{S}$ in descending order as: $F(\bm{x}^{[1]})\geq F(\bm{x}^{[2]})\geq ... \geq F(\bm{x}^{[S]})$\;

    Select elites as $\bm{x}^{[1]}$, $\bm{x}^{[2]}$, ..., $\bm{x}^{[S_\text{elite}]}$\;

    Update $\bm{u}^{(t+1)}$ according to (\ref{eq:smooth_update})\;

    $t=t+1$\;

    \textbf{end for}\\
    \textbf{Output:} $\bm{x}^{[1]}$ as the optimal association.
    \caption{The CE-based ASsociation (CEAS)}
\end{algorithm}
\DecMargin{1em}
For simplification, we assume a single-antenna system model. Joint optimization of association and beamforming design for multiple-antenna system is in general much more involved \cite{Shen2014Distributed}. To extend the proposed algorithm to multi-antenna scenarios, an alternating optimization mechanism can be adopted \cite{Shen2014Distributed}. We can firstly conduct the association optimization in an equivalent SISO network by the proposed CEAS algorithm. Beamforming optimization with given association can thus be conducted with existing methods.

\subsection{Complexity Analysis}
\vspace{-0.1cm}
In \textbf{Algorithm 1}, the main complexity of CEAS obviously comes from steps 4 and 7. In step 4, the objective function of each candidate is calculated, which involves the complexity order of $\mathcal{O}(S)$. In step 7, the probability parameter needs to be updated according to (\ref{eq:smooth_update}) with the complexity $\mathcal{O}(N)$. Therefore, the total complexity of CEAS amounts to $\mathcal{O}(T(S\!+\!N))$. Note that the choice for $S$ depends on the size of the problem and it is suggested to take $S\!=\!cN$, where $c$ is a constant \cite{Boer2005A}. Thus the complexity of CEAS amounts to $\mathcal{O}(N)$. Table \ref{TABLE:complexity} lists the complexity of different algorithms. From Table \ref{TABLE:complexity}, we conclude that the computational complexity of the proposed CEAS algorithm is comparable to existing convex optimization methods and significantly lower than the exhaustive search.

Besides, unlike convex optimization algorithms, CEAS is nonsensitive to parameter choices and network deployment, which is convenient to adjust adaptively in dynamic HetNets. Note that the proposed CEAS algorithm runs without any prior information and achieves nearly optimal performance as illustrated in the next section. In practice, some prior information, e.g., previous associations in the network, are available, which can be utilized to accelerate convergence and further reduce complexity.
\begin{table}[t!]
  \setlength{\belowcaptionskip}{-10pt}
  \centering
  \footnotesize
  \caption{Complexity Comparison of Different Algorithms}\label{TABLE:complexity}
  \begin{tabular}{|c|c|c|c|}\hline
    & CEAS & Dual-based [2] & Exhaustive search \\ \hline
  Complexity & $\mathcal{O}(N)$  & $\mathcal{O}(N)$ & $\mathcal{O}(2^{N})$ \\ \hline
\end{tabular}
\end{table}

\section{Numerical Results}\label{sec:simulation}
\vspace{-0.1cm}
In this section, we evaluate the performance of the proposed CEAS algorithm via simulation. We compare CEAS algorithm with the existing association methods, i.e., the Max-SINR association and the convex optimization algorithm in \cite{Ye2012User} based on Lagrangian dual decomposition. Consider a downlink 2-tier HetNet with one MBS and three SBSs per cell. The transmission powers of MBS and SBSs are $\{43, 23\}$ dBm. Thirty users are uniformly distributed in a cell with radius 500 m. The system bandwidth is $W=10$ MHz, and the path loss is modelled as $128.1+37.6\log_{10}d$(km).

\begin{figure}[t!]
\setlength{\abovecaptionskip}{-4pt}
\centering
\includegraphics[width = 3.2in]{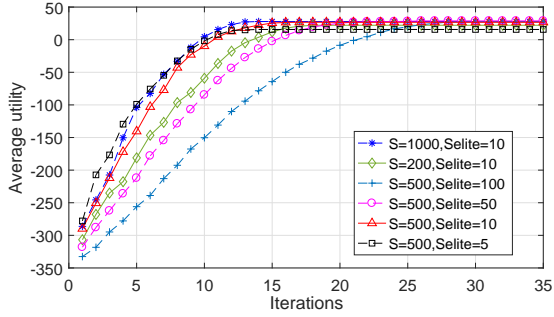}
\caption{Average utility against iterations with varying $S$ and $S_\text{elite}$.}
\label{Fig:varyS}
\vspace{-0.3cm}
\end{figure}
Fig. \ref{Fig:varyS} shows the average utility value against the number of iterations under different values of $S$ and $S_\text{elite}$. From Fig. \ref{Fig:varyS}, it is observed that increasing $S$ appropriately is beneficial for both accelerating the convergence and improving the objective value. However, when $S$ becomes sufficiently large, e.g., $S\!=\!500$, such benefits disappear. In contrast, the impact of $S_\text{elite}$ is quite different. When $S_\text{elite}$ is large, decreasing $S_\text{elite}$ makes the algorithm converge quickly while still achieves the same performance. However, if $S_\text{elite}$ is too small, the convergence speed is accelerated at a cost of performance degradation. Therefore, the values of $S$ and $S_\text{elite}$ can be adjusted for balancing the complexity and performance. In our algorithm, we simply set $S\!=\!500$, $S_\text{elite}\!=\!10$, and $T\!=\!20$.

\begin{figure}[t!]
\setlength{\abovecaptionskip}{-4pt}
\centering
\includegraphics[width = 3.2in]{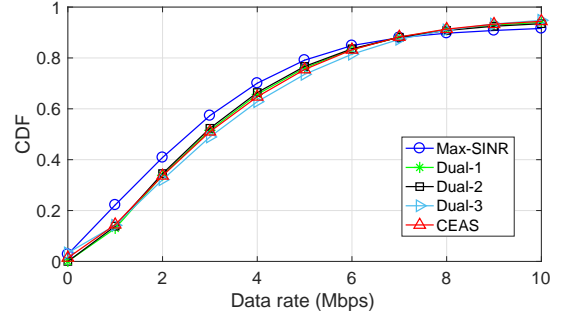}
\caption{CDF of data rates under different schemes.}
\label{Fig:CDF}
\vspace{-0.3cm}
\end{figure}

\begin{figure}[t!]
\setlength{\abovecaptionskip}{-4pt}
\centering
\includegraphics[width = 3.2in]{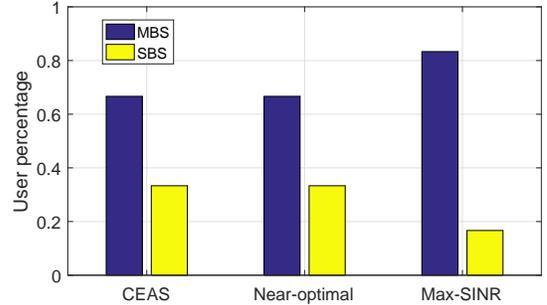}
\caption{The percentage of MBS/SBS users for different association methods.}
\label{Fig:user percentage}
\vspace{-0.3cm}
\end{figure}

\begin{table}[t!]
  \setlength{\belowcaptionskip}{-10pt}
  \centering
  \footnotesize
  \caption{Average Utility and Rate Comparison of Different Algorithms}\label{TABLE:utility}
  \begin{tabular}{|c|c|c|c|c|c|}\hline
    & CEAS  & Max-SINR  & Dual-1 & Dual-2 & Dual-3 \\ \hline
UEE & 37.153 & 32.575 & 37.105 & 36.679 & 35.684 \\ \hline
Rate (Mbps) & 4.7632 & 5.0843 & 4.7394 & 4.6807 & 4.6535  \\ \hline
\end{tabular}
\end{table}
For performance comparison, we test the association algorithm in \cite{Ye2012User} with different parameters, which are referred to as ``Dual-1'', ``Dual-2'' and ``Dual-3'' in Fig. \ref{Fig:CDF}. Fig. \ref{Fig:CDF} plots the cumulative distribution function (CDF) of data rates. Table \ref{TABLE:utility} lists the average utility and rate obtained by these methods. From Fig. \ref{Fig:CDF} and Table \ref{TABLE:utility}, we find that the performance of the convex optimization algorithm in \cite{Ye2012User} obviously varies with the parameter choices for the dual-based algorithm. In contrast, our proposed CEAS algorithm is shown to outperform ``Max-SINR'' association by 14\% in terms of utility and achieves the best utility among all comparison methods. Note that although the ``Max-SINR'' achieves the highest average rate, it results in extremely unfair user experience as evidenced. Fig. \ref{Fig:user percentage} compares the percentage of MBS/SBS users for different association methods. ``Near-optimal'' denotes the algorithm in \cite{Ye2012User} with proper parameters. It shows that the proposed CEAS algorithm achieves better load balancing while the ``Max-SINR'' association results in the overload of MBS.

\section{Conclusion}\label{sec:conclusion}
\vspace{-0.1cm}
In this letter, we considered the user association in HetNet and formulated a utility maximization problem aiming to achieve load balancing. We proposed an alternative association algorithm by employing the idea of CE optimization in machine learning, which is robust to network deployment and algorithm parameter choices. Results demonstrated that our proposed CEAS algorithm achieves near-optimal performance efficiently. The user association problem with explicit individual user QoS constraints is a future research of interest.


\begin{thebibliography}{1}
\footnotesize

\bibitem{Yang2017User}
Z. Yang \emph{et al.}, ``User association, resource allocation and power control in load-coupled heterogeneous networks,'' in \emph{Proc. IEEE Global Commun. Conf. Workshops}, Washington, DC, USA, Feb. 2017, pp. 1--7.

\bibitem{Ye2012User}
Q. Ye \emph{et al.}, ``User association for load balancing in heterogeneous cellular networks,'' \emph{IEEE Trans. Wireless Commun.}, vol. 12, no. 6, pp. 2706--2716, June 2013.

\bibitem{Huang2018Utility}
X. Huang \emph{et al.}, ``Utility-energy efficiency oriented user association with power control in heterogeneous networks,'' \emph{IEEE Wireless Commun. Lett.}, accepted to appear, early access Jan. 2018.

\bibitem{Shen2014Distributed}
K. Shen and W. Yu, ``Distributed pricing-based user association for downlink heterogeneous cellular networks,'' \emph{IEEE J. Sel. Areas Commun.}, vol. 32, no. 6, pp. 1100--1113, June 2014.


\bibitem{Rubinstein2004The}
R. Y. Rubinstein and D. P. Kroese, \emph{The Cross Entropy Method: A Unified Approach to Combinatorial Optimization, Monte-carlo Simulation (Information Science and Statistics).} Secaucus, NJ, USA: Springer-Verlag New York, Inc., 2004.

\bibitem{Kovaleva2017Cross}
M. Kovaleva \emph{et al.}, ``Cross-entropy method for electromagnetic optimization with constraints and mixed variables,'' \emph{IEEE Trans. Antennas Propag.}, vol. 65, no. 10, pp. 5532--5540, Oct. 2017.

\bibitem{Han2016A}
T. Han and N. Ansari, ``A traffic load balancing framework for software-defined radio access networks powered by hybrid energy sources,'' \emph{IEEE/ACM Trans. Network.}, vol. 24, no. 2, pp. 1038--1051, Apr. 2016.

\bibitem{Han2017Backhaul}
Q. Han \emph{et al.}, ``Backhaul-aware user association and resource allocation for energy-constrained HetNets,'' \emph{IEEE Trans. Veh. Technol.}, vol. 66, no. 1, pp. 580--593, Jan. 2017.

\bibitem{Liu2016User}
D. Liu \emph{et al.}, ``User association in 5G networks: A survey and an outlook,'' \emph{IEEE Commun. Surveys \& Tutorials}, vol. 18, no. 2, pp. 1018--1044, 2016.

\bibitem{Rubinstein2013The}
R. Y. Rubinstein and D. P. Kroese, \emph{The Cross-entropy Method: A Unified Approach to Combinatorial Optimization, Monte-carlo Simulation and Machine Learning.} Springer Science \& Business Media, 2013.

\bibitem{MacKay2003Information}
D. MacKay, \emph{Information Theory, Inference, and Learning Algorithms.} Cambridge University Press, 2003.

\bibitem{Boer2005A}
P. D. Boer \emph{et al.}, ``A tutorial on the cross-entropy method,'' \emph{Annals of Operations Research}, vol. 134, no. 1, pp. 19--67, 2005.

\end{thebibliography}
\end{document}